# A Study of User's Performance and Satisfaction on the Web Based Photo Annotation with Speech Interaction

Siti Azura Ramlan, and Nor Azman Ismail

**Abstract**— This paper reports on empirical evaluation study of users' performance and satisfaction with prototype of Web Based speech photo annotation with speech interaction. Participants involved consist of Johor Bahru citizens from various background. They have completed two parts of annotation task; part A involving PhotoASys; photo annotation system with proposed speech interaction and part B involving Microsoft Microsoft Vista Speech Interaction style. They have completed eight tasks for each part including system login and selection of album and photos. Users' performance was recorded using computer screen recording software. Data were captured on the task completion time and subjective satisfaction. Participants need to complete a questionnaire on the subjective satisfaction when the task was completed. The performance data show the comparison between proposed speech interaction and Microsoft Vista Speech interaction applied in photo annotation system, PhotoASys. On average, the reduction in annotation performance time due to using proposed speech interaction style was 64.72% rather than using speech interaction Microsoft Vista style. Data analysis were showed in different statistical significant in annotation performance and subjective satisfaction for both styles of interaction. These results could be used for the next design in related software which involves personal belonging management.

——————————   ◆   ——————————

## 1 INTRODUCTION

People at present time possess large collections of digital photos caused by digital photo technology that allows them to produce photos easily. In transporting personal digital photos, InfoTrends/CAP Ventures has conducted a survey and reported that over 825 million photos are stored at online photo services, along with sharing and printing activity which continue to increase as well as uploading activity involvement. It has encouraged the development of digital photos systems to provide the digital photos management using Internet platform. Recently, many photo management system software are developed either in stand-alone system or Web-based system such as iPhoto, Flickr, Picasa online, MyPhotoAlbum, and Snapfish. The latest software which is similar to the PhotoASys is the stand-alone iPhoto'09 invented by MacApple which uses 4w's (where, when, who and what) idea.

Multimodal interfaces allow multiple mode of interaction such as speech, voice, stylus, gesture, eye tracking etc. Many researchers have done and currently investigating on multimodal interface area for photo annotation. For instance, Chen and Tan in 2001 and 2003 developed an annotation system using speech. Rohini et al. also used speech in Show&Tell system. For the meantime, the advantages of using speech-activated commands over mouse-keyboard-activated commands have been demonstrated by Karl et al. From the research, the reduction in task time due to using speech was 18.67%. The efficiency of input mode in multimodal dialogue systems has been investigated by M. Perakakis et al.

Useful annotation process method for digital photos collection is necessary to ease photo management especially when users want to search or browse through their photos. Table 1 shows three types of annotation techniques in human effort and machine assistance contextual. Annotation process on both digital photos and traditional printed photos are not easy, especially for thousands of photo collections.

Many researchers have tried to find that best way to make annotation process and causing existing variety of the annotation techniques and styles which have been produced in previous researches. For example, EasyAlbum was introduced as a method of annotating photograph based on face clustering and re-ranking. It uses 'who' category to annotate the photos collection. B. Suh et al chose 'what and 'who' category which discusses on annotating photos based on hierarchical events clustering or multiple levels of event groups and torso based human identification. Torso based human identification was chosen due to low accuracy of human face recognition. Nevertheless, this system fails to work properly for photos containing people wearing clothes or uniforms that look similar to one another. In indexing the photos, J. Chen et al, defined four fields namely people, place, date and event that can be implemented in annotating photos. However they used Nuance Dragon Naturally Speaking to perform the speech interaction in

————————————————

- *Siti Azura Ramlan is the postgraduate student of Faculty of Computer Science and Information System, Universiti Teknologi Malaysia*
- *Nor Azman Ismail a senior lecturer recently working in Faculty of Computer Science and Information System, Universiti Teknologi Malaysia.*

the annotation system. Meanwhile, K. Christian et. al have conducted their study and presented that navigation by voice control increases the time performance for certain task. Therefore, annotation is a wide issue to be discussed which relates to the retrieval part of the process pattern. A high technology camera will open a chance for users to ignore annotation due to time (when) and location (where) – latest technology using GPS with automatic annotation) automatically created when the photo is taken. It will immediately process the annotation for when and where category. However, how well the users will remember the details of photos taken?

TABLE 1
ANNOTATION TECHNIQUE IN HUMAN EFFORT AND MACHINE ASSISTANCE CONTEXTUAL.

|  | **Human Effort** | **Machine Assistance** |
|---|---|---|
| **Manual** | Add structured annotations, with sufficient semantic information to be useful for retrieval. | Save annotations in a database. |
| **Semi-Automatic** | Add freestyle annotations or captions. Potentially work with machine's output in an iterative fashion. | Parse human-entered captions and extract semantic information. |
| **Automatic** | Verify machine's accuracy and make corrections as needed. | Add structured annotations using GPS, context, or recognition technology. |

Additionally, textual annotation is time-consuming and cost-consuming. It requires users to type a word one by one and that certainly take a long time to complete the annotation for all photo collections. Therefore, the subjectivity of annotation should be compressed to solve all the obstacles mentioned above. So, the system of annotation was designed to be applied in photo annotation system. This system, PhotoASys applied 4w's metadata category to make systematic annotation system; when, what, who and where. Figure 1 depicts the screen shot of PhotoASys. This paper will report on an empirical evaluation of users' performance and satisfaction with Web based speech photo annotation as well as comparison in statistical significance between proposed speech interaction and Microsoft Vista speech interaction.

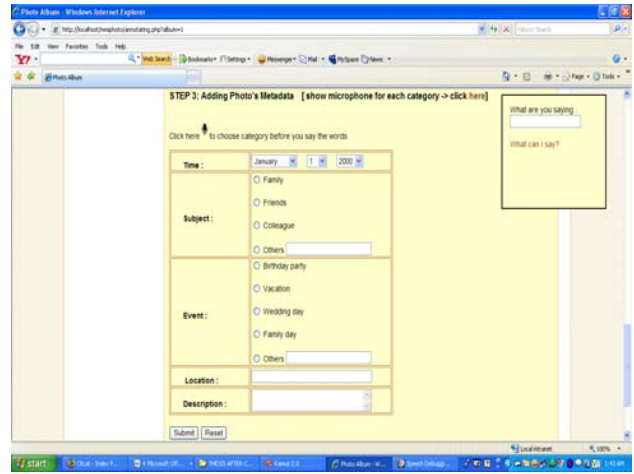

Fig 1. Screen-shot of PhotoASys

## 2 SYSTEM OVERVIEW

PhotoASys is a prototype of the personal photo annotation with enabled speech interface. This prototype system is based on World Wide Web environment and it was developed for two styles of interaction; PhotoASys with proposed speech interaction and PhotoASys with Microsoft Vista speech interaction. This study is meant to get the difference between both styles of the speech interface in which the proposed speech interaction was developed (refer to figure 1). Speech with Microsoft Vista was developed by Microsoft for general application that enables speech by using the Microsoft Vista speech recognition. Users need to annotate photos for easier retrieval and are allowed to annotate with four categories of photo, photo metadata (when, what, who and where). Therefore, users are given an opportunity to use speech as an alternative way to enter the input as compared to the traditional way which uses keyboard, traditional way.

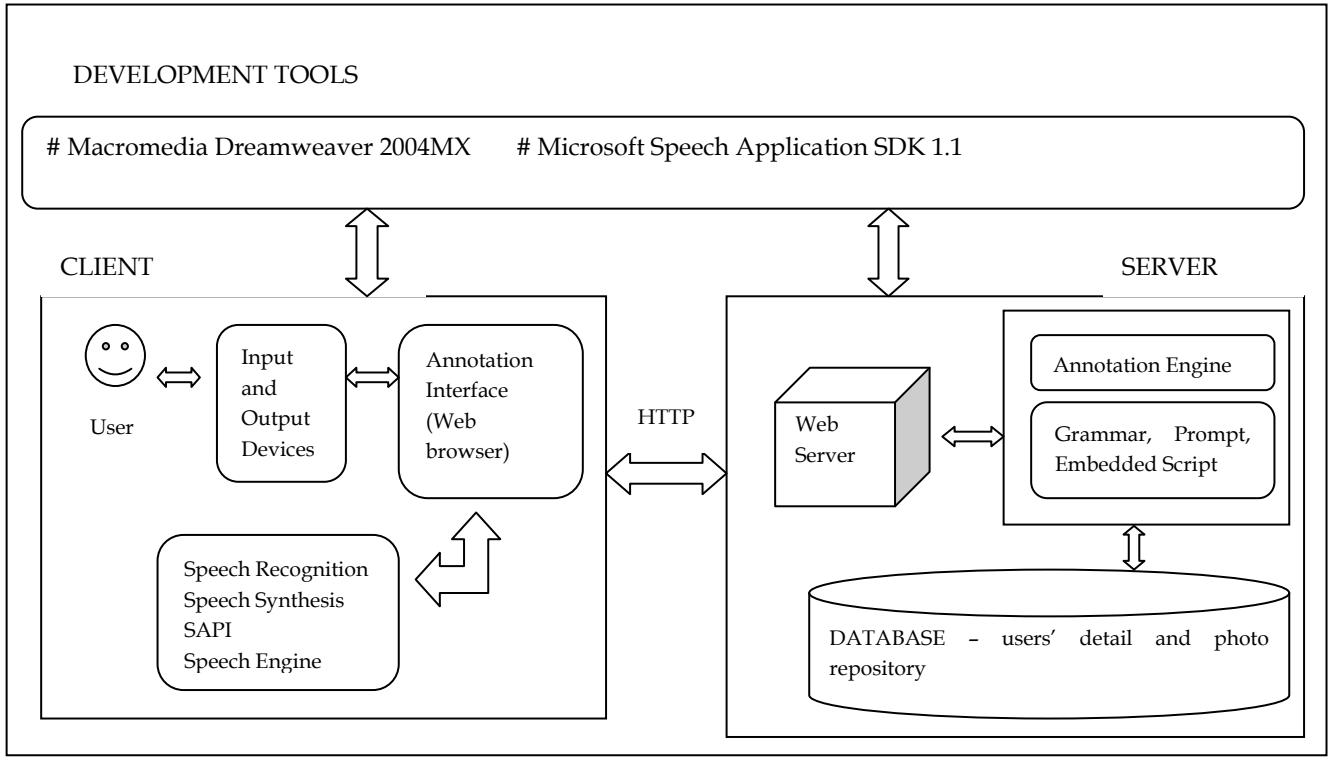

Fig. 2. System Architecture of PhotoASys

Figure 2 shows the architecture of the PhotoASys with development tools which was designed for proposed speech interaction style and developed using Speech Application SDK version 1.1 from Microsoft. The architecture consists of client and server with development tools. This architecture enables users to annotate personal digital photo by four categories of photo metadata; time (when), subject (who), event (what) and place (where). The data have been saved in the photo repository by using both speech style interactions. The system is not limited to annotate photo only, users are also allowed to navigate the interface system such selecting menu, selecting album, selecting photo and data submission. All users' details and photos' repository are stored in the database in web server. The other components in web server are scripting program, grammar, prompt and annotation engine. In the client side, Internet Explorer is used to run the prototype system.

The prototype system with proposed speech interaction was developed using several web technologies. It was built using web programming scripting language Javascript, XHTML, PHP, XML mark-up language (Speech Application Language Tags, SALT). For PhotoASys with Microsoft Vista speech interaction style, the prototype system would publish in Windows Microsoft Vista platform and activates the speech in setting part of speech recognition since Microsoft Vista has built in speech recognition.

## 2.1 Annotation Commands and Dialogue Management

PhotoASys with proposed speech interaction style

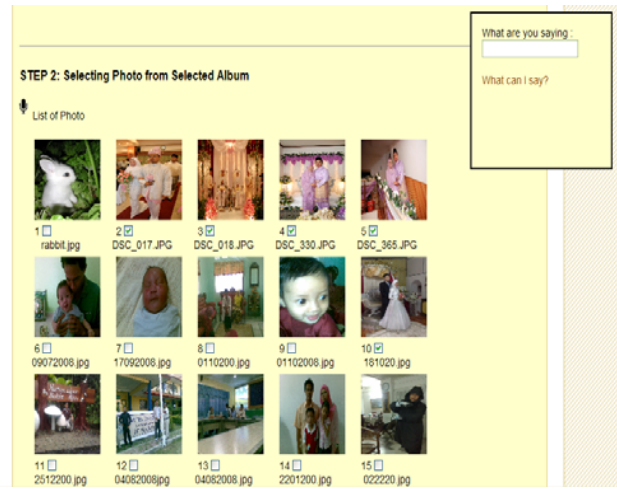

Fig. 3(a). photoASys user interface

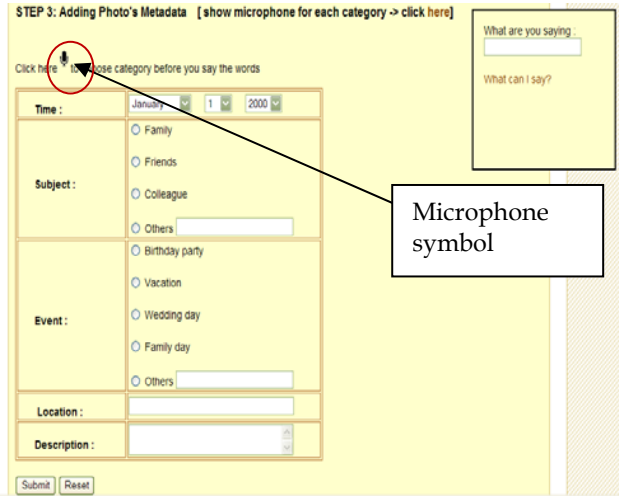

Fig. 3 (b). PhotoASys user interface

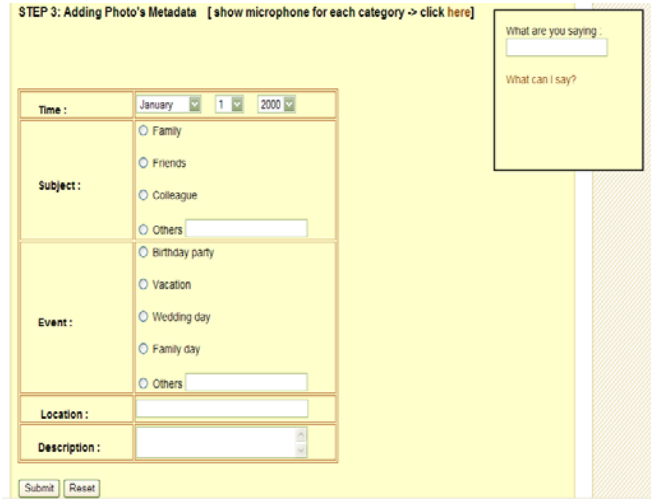

Fig 4 (b). PhotoASys user interface

Figure 3(a) and 3(b) show the user interface of PhotoASys while 3(a) shows the selecting photo part. For PhotoASys with proposed speech interaction style, users need to click the activate button (microphone symbol: in red circle) then say the instruction either to select (check) or delete (uncheck) followed by the number of photo. Users can choose the photos after successful login and further annotate the photo with entered input in 3(b) part. The four categories of photo metadata are 4w's (when (time), who (subject), what (event) and where (location) category) and an additional description category.

PhotoASys with Microsoft Vista speech interaction style

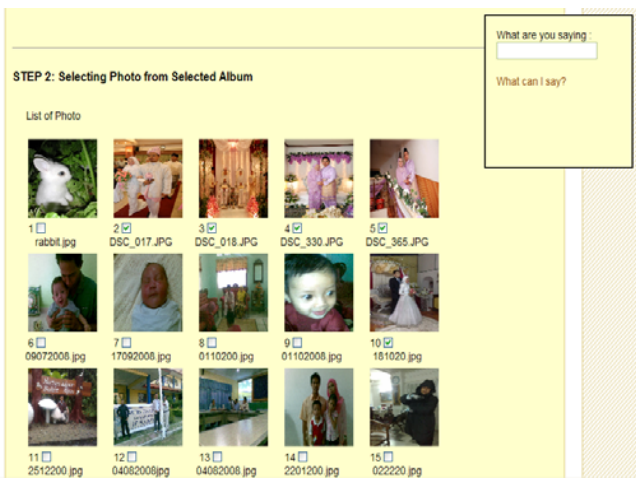

Fig. 4(a). PhotoASys user interface

Figure 4 displays the user interface of PhotoASys with Microsoft Vista speech interaction style. This prototype system is a personal digital photo annotation system based on Web and embedded with the Microsoft Vista recognition system that has built in Microsoft Vista Operating system platform. The speech in this system should follow Microsoft Vista Speech Interaction manual so that the system can hear input/instruction from users.

## 3 METHODOLOGY

### 3.1 Research Design

Participants involved consist of Johor Bahru citizens from various background of computer using. The experiment was conducted in a modified lab which has been appropriately setup for the experiment. Participants could only come once for the experiment. Questions were asked prior they performed tasks give. The questionnaires include computer using experience, annotation background knowledge and web-based background knowledge.

At the beginning of the experiment, they were given description of the experimental purpose and procedure which they need to follow throughout the experiment process. Then, they were given 15 minutes for exploration so that they could be familiar with the terms and ways to interact with the system. The system requires the participants to interact by voice. Therefore, participants were needed to go through voice training profile in speech properties in control panel before the experiment began. The objective for this process is to make the computers recognize the speech pattern and become familiar with the voice. Then, the voice will be selected from the speech as input modalities for the annotation system.

All participants were then given the annotation task which consists of two main parts. First part necessitates the participants to complete the task by using the proposed speech interaction style in which participant will click the activate button and speak the right word to give input to the system. Second part is system with Microsoft Vista Speech Interaction style. Every main part consists eight sub tasks that should be completed by the participants. The participants were told that they were given 3 minutes to complete each task. The experiment was stopped when the participant did not manage to complete task in given period of time. The participants were allowed to move to other task even though they could not complete the previous task. They were also given help menu in the system for assistance.

Twenty-two volunteers were offered to take part in the experiment. In average, they have five years experience in computer using. All of them owned digital photos collection. In annotation experience, nine of total participant knew hoe to annotate by keyword. All participants knew how to use Internet Explorer.

### 3.2 Annotation Task

All participants were given two main parts of the task and each task has eight sub tasks to be completed. The whole tasks are as follows:

Task 1: Login into the system.
Task 2: Select the given album.
Task 3: Select the photo(s).
Task4: Enter the input for time category.
Task 5: Enter the input for subject category.
Task 6: Enter the input for event category.
Task 7: Enter the input for location category.
Task 8: Enter the input for description category.

Task 1, task 7 and task 8 represented in text field which is what the input says will be converted to text in the text field. Task 2 is selecting album that will be represented in slide show of album. Figure 5 shows the screen shot on selecting photos. The checkbox will become active when the users say the number of photo that they have selected and uncheck when users give the delete command. Radio button will check when the users have chosen the selection for subject and event meanwhile drop down menu will change the input of time that users have selected. This means that participants need to complete 16 tasks in order to complete both parts with different styles of interaction.

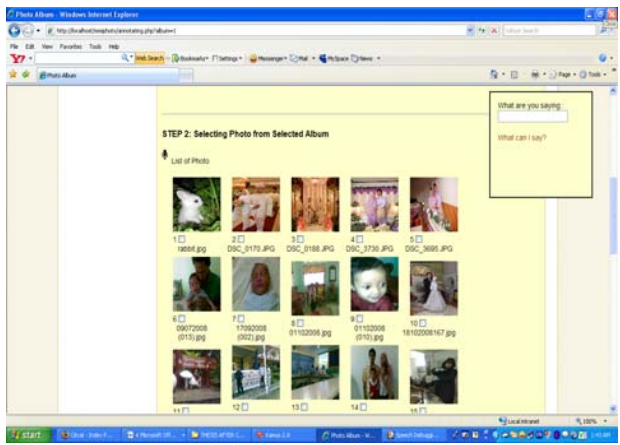

Fig. 5. Screen-shot of selecting photos

### 3.2 Data Collection

Two methods were used to collect the data namely screen recording and questionnaires. The screen recording begins when participant starts to perform the experiment. The Camtasia Studio v6.0 software was used to record the screen during the experiment. The screen recording was be used to note the time taken.

After completing all annotation tasks for screen recording, participants need to complete questionnaires about the annotation task for both parts and overall reaction of the system. The questionnaires were designed based on Questionnaire for User Interface Satisfaction (QUIS) and previous research by S. M Zabed Ahmed [20] who has conducted a study in Human-Computer Interaction. Participants were needed to rate their satisfaction in this method either positive or negative for each aspect in the prototype system.

In this study, the data collection was analyzed according to two types of variables; the independent variables and dependent variables. The independent variables in this study are as follows:

- PhotoASys with proposed speech interaction, $M_P$
- PhotoASys with Microsoft Vista Speech interaction, $M_v$.

Which PhotoASys is a photo annotation system. Both variables above were then analyzed to get the level of statistic significant for both styles. The dependent variables used in the study are as follows:

Time completion: Time taken to complete the task.

Subjective Satisfaction: Measured by questionnaires given to participants after completion of the experiment. All responses data have analyzed the prototype user interface.

First and second dependent variables were collected from the screen recording along the experiment. Meanwhile, the second variable of data was obtained from the completed. The other variable has also been analyzed including an acceptability of the system collected from the questionnaires.

From the variables studied and data collection, the null hypotheses for this study are listed as the following:

- Hypothesis 1: No difference between using prototype with proposed speech interaction style and using prototype with Microsoft Vista speech interaction in user annotation performance.
- Hypothesis 2: No difference between using prototype with proposed speech interaction style and using prototype with Microsoft Vista speech interaction in subjective satisfaction.

## 4 RESULTS

### 4.1 Time Completion

The time completion has been collected from video recording of the evaluation session. The time duration of task was taken in minutes. Table 2 shows the mean and standard deviation of time completion of annotation task with proposed speech interaction style and with Microsoft Vista Speech interaction. Mean for proposed speech interaction style is represented as $M_c$ while mean for Microsoft Vista Speech interaction is represented as $M_v$. Then, $SD_P$ and $SD_v$ represent Standard Deviation value for each proposed speech interaction and Microsoft Vista Speech interaction. From the table, tasks 4 until task 7 are the annotation task for 4w's category. For these tasks, task 4; time category noted the highest time taken to complete the task with $M_P$ = 0.415 as compared to the other three categories by using proposed speech interaction style. This is due to more input in time category namely month, day and year. Even though, time category for proposed speech interaction style took more time than other categories, time category in Microsoft Vista speech interaction style took even longer time to complete the task, $M_v$ is 1.312. This is due to the fact that users were needed to go through two steps to enter an input. There were six steps to complete annotation task for time category. Subject and event category noted time taken with Mp = 0.105 and 0.084 for each category and Mv = 0.191 and 0.161 for each category. Time taken for both interaction styles was not vastly different because it involves two steps for proposed speech interaction style and three steps for Microsoft Vista speech interaction style. Meanwhile, place or location category took 0.117 for Mp and 0.3380 for Mv whereas two steps were involving in proposed speech interaction style and three steps for another style. Thus, all evaluation results are more dependent on the dialogue flow (dialogue system).

Next, the total mean of time taken for $M_P$, 1.807 is less than $M_v$ with 4.781. The value of Standard Deviation also shows that $SD_P$ is less than $SD_v$. The table indicates the value of Standard deviation for proposed speech interaction is 1.822 and 2.875 for Microsoft Vista Speech interaction style. This is because Microsoft Vista Speech interaction needs more steps to complete a task rather than proposed speech interaction style. Microsoft Vista Speech interaction style was designed for all application and program that run in Microsoft Microsoft Vista platform. In contrast, proposed speech interaction was specially designed for Web Based Photo Annotation and unable to run in other systems. Hence, it is different between application designed for general and specific program. Therefore, Microsoft Vista speech interaction is too flexible to be applied to general system and it causes users to take longer time to learn the system. Therefore, statistical technique was used to see the difference between statistical significance of proposed speech interaction and Microsoft Vista Speech interaction. However, the normality distribution of time taken value was needed to determine the hypotheses would be accepted or rejected.

From the normality test, the sig. value of the Kolmogorov-Smirnov statistic was 0.021 and 0.200 for each proposed style of speech interaction. From the normality test, the sig. value of the Kolmogorov-Smirnov statistic was 0.000 and 0.200 for each proposed speech interaction and Microsoft Vista Speech interaction style. This means that non-normality in which the Sig. value of proposed speech interaction was less than 0.05. It might be because of variation of data collection due to various backgrounds of participants involved.

From the normality test, Wilcoxon Signed Rank test was used to see the difference between proposed speech interaction style and Microsoft Vista Speech interaction style for Web Based photo annotation in time completion. The summary of the results are shown in table 3. The table shows that there is a significant difference in time completion between proposed speech interaction style and Microsoft Vista Speech interaction style for Web Based photo annotation. The null hypotheses (H1) have been rejected due to significance level of 0.000 which is less than 0.05.

TABLE 2
MEANS AND STANDARD DEVIATIONS OF TIME COMPLETION OF ANNOTATION TASK

| Task | Time taken (mins) | | | |
|---|---|---|---|---|
| | $M_P$ | $M_v$ | $SD_P$ | $SD_v$ |
| task1 | 0.370 | 1.184 | 0.536 | 0.543 |
| task2 | 0.241 | 0.769 | 0.180 | 0.576 |
| task3 | 0.309 | 0.498 | 0.402 | 0.471 |
| task4 | 0.415 | 1.312 | 0.375 | 0.622 |
| task5 | 0.105 | 0.191 | 0.065 | 0.203 |
| task6 | 0.084 | 0.161 | 0.050 | 0.050 |
| task7 | 0.117 | 0.380 | 0.106 | 0.188 |
| task8 | 0.166 | 0.286 | 0.108 | 0.222 |
| Total | 1.807 | 4.781 | 1.822 | 2.875 |

*$M_P$: Mean for proposed speech interaction style
$M_V$: Mean for Microsoft Speech interaction style
$SD_P$: Standard Deviation for proposed speech interaction style
$SD_V$: Standard deviation for Microsoft Vista speech interaction style

TABLE 3
WILCOXON SIGNED RANK TEST FOR TIME COMPLETION WHEN USING DIFFERENT STYLE OF INTERACTION.

| | | N | Mean Rank | Sum of Ranks | Z | Asymp. Sig. (2-tailed) |
|---|---|---|---|---|---|---|
| Total time vista speech interaction – Total time proposed speech interaction | Negative Ranks | 2[a] | 2.50 | 5.00 | -3.945[a] | 0.000 |
| | Positive Ranks | 20[b] | 12.40 | 248.00 | | |
| | Ties | 0[c] | | | | |
| | Total | 22 | | | | |

a. total time vista speech interaction < total time proposed speech interaction

b. total time vista speech interaction > total time proposed speech interaction

c. total time vista speech interaction = total time proposed speech interaction

## 4.2 Subjective Satisfaction

The subjective satisfaction data collection was collected from the questionnaires that participants need to complete after the screen recording session of the experiment. They were given two sets of questionnaire; part A to show feedback of the proposed speech interaction style and part B to give feedback of Microsoft Vista Speech interaction style which they have experienced through the experiment. They were given score as a feedback to the prototype system.

The Cronbach's Alpha coefficient is commonly used to demonstrate the reliability of the question and scale. The reliability is considered reliable when the value is more then 0.7. So, the Cronbach's Alpha has been used to assess the reliability of the questions. The value of α = 0.945 indicates that subjective satisfaction when using proposed speech interaction style was highly reliable. Meanwhile, Microsoft Vista speech interaction also shows high reliability with the α = 0.971.

The Likert-type scale with nine points was used to measure the subjective satisfaction and was rated by all participants. Score 1 indicates negative perception (low optimism) while score 9 demonstrates positive perception (high optimism). The subjective satisfaction questionnaires have 12 questions for every part. The question could be grouped into a few parts namely Q1 to Q6 to rate the overall reaction of the system, Q7 and Q8 to measure the satisfaction of Login part and help components as well as Q9 until Q12 to indicate the satisfaction of annotation task reaction. Table 4 shows the average subjective satisfaction scores for proposed speech interaction style and Microsoft Vista Speech interaction style.

TABLE 4
AVERAGE SCORE FOR BOTH STYLES OF INTERACTION IN SUBJECTIVE SATISFACTION

| Question | | Proposed speech interaction | Microsoft Vista speech interaction |
|---|---|---|---|
| **Overall reaction** | | | |
| Q1 | Wonderful | 6.818 | 5.182 |
| Q2 | Satisfying | 6.682 | 5.227 |
| Q3 | Stimulating | 6.455 | 5.364 |
| Q4 | Easy | 6.273 | 4.818 |
| Q5 | Adequate power | 6.455 | 4.909 |
| Q6 | Flexible | 6.091 | 4.682 |
| **Login and understanding help** | | | |
| Q7 | Login into the system: | 6.455 | 4.864 |
| Q8 | Help components: | 6.682 | 5.409 |
| **Annotation tasks** | | | |
| Q9 | Annotate photo in when (time) category | 7.182 | 6.045 |
| Q10 | Annotate photo in what (event) category | 7.227 | 5.818 |
| Q11 | Annotate photo in who (subject) category | 7.273 | 5.727 |
| Q12 | Annotate photo in where (location) category | 7.182 | 5.591 |

TABLE 5
PAIRED SAMPLE T-TEST FOR SUBJECTIVE SATISFACTION USING DIFFERENT STYLE OF INTERACTION

| | Tap and Talk Subjective Satisfaction<br><br>Mean<br>Std. Deviation | Microsoft Vista Speech recognition Subjective Satisfaction<br><br>Mean<br>Std. Deviation | t-value | df | Sig. (2-tailed) |
|---|---|---|---|---|---|
| Subjective Satisfaction for different style of interaction | 80.77<br>13.777 | 63.64<br>17.751 | -4.856 | 21 | 0.000 |

Table 4 and 5 show that collection scores for proposed speech interaction style is more than collection score for Microsoft Vista speech interaction style. This means that users are more comfortable with proposed speech interaction although it is not too flexible as compared to Microsoft Vista speech interaction. The score was not very high (not almost score 9) since users come from various background and most of them do not have any experience in using speech style of interaction. They still can understand the system despite being novice users. Further statistical analysis was conducted to determine the significant difference in subjective satisfaction when using different style of interaction. From the normality test, the Kolmogorov-Smirnov statistic shows the Sig. value was 0.200 for proposed speech and Microsoft Vista speech style of interaction. It is a normal score since Sig. value of proposed speech interaction was more than 0.05. From the results of normality testing, the suitable technique to test the difference of statistical significant is paired samples t-test. Table 5 shows that there was significant difference in subjective satisfaction between proposed speech and Microsoft Vista Speech style of interaction for Web Based photo annotation. The null hypothesis (H2) is rejected since significance level was 0.000 which is less than 0.05.

## 5 DISCUSSION

Overall evaluation shows that proposed speech interaction style is more appropriate for Web-based photo annotation system, PhotoASys rather than Microsoft Vista speech interaction style. Table in previous section shows that proposed speech interaction style gives more subjective satisfaction to users in short period of time (refer to table 2 and table 4).

Dialogue system used provides more impact to the evaluation and data analysis because both of interaction styles are not related to the Fitts' law. Fitts' law usually relates to the Website interface design where any movement towards the target should be precise and quick in pointing the target. It depends on various factors such size of object, position of object, shape of object etc. In this system, Fitts' law was not involved in the evaluation session since the evaluation only analyzed for the speech interaction style. Speech dialogue does not depend on the Fitts' law. Therefore, this system was closed and speech dialogue system was developed.

Data analysis involved two types of speech interaction style, called variable study. There are proposed speech interaction style and Microsoft Vista speech interaction style. The different of both interaction style is the utilization of recognition mode. Proposed speech interaction style used single mode speech interaction while multimode interaction was applied in the PhotoASys with Microsoft Vista speech interaction style. Users took a short time to complete all tasks when they used proposed speech interaction style due to the easy concept of dialogue system. All tasks have similar style, it is "tap and talk". Users just need to click the activate button and say the input or instruction when the wave signal appeared. Single mode recognition has time period to enter the input. Thereby, users know when they should say the input and the input would recognize according to the button that users activated before. The system knows which input it is. Therefore, the proposed speech interaction style was neither too rigid nor too flexible. It is sufficient with what the users want and need. In contrast, Microsoft Vista speech interaction style took a longer time to complete all tasks. It used multimode recognition that used open-microphone concept which does not have period time to enter the input. Users need time to learn the flow of dialogue due to many instruction and alternative ways to complete the task. It has caused confusion to the users. For example, user can say the clickable button or by using special word, "show number".

Single mode recognition has minimum rate of noise interruption caused by time period in which this recognition mode supported recognition of voice. System only listens when the activation button was clicked until the wave signal disappeared. It is very helpful for users to say the input. Besides, this proposed speech interaction style allowed the integration of speech synthesis where users could hear the feedback or instruction for correction if any error has occurred. From the feedback, users would know the next action that they need to do. Speech synthesis complement the dialogue flow developed and allowed response after any errors such as silence event, no recognition event etc.

The Microsoft Vista speech interaction style was applied using open-microphone concept which return the results at intervals along the speech detection time period and always ready to receive any input from users as long as microphone is opened. This has caused easy identification of noise interruption. It does have speech synthesis to support the feedback and response by voice for any error or silent event. It also has response in microphone panel that users need to read and follow the next instruction. Besides of surrounding interruption, pronunciation also gives an impact to the input especially input for text field. In contrast for navigation, Microsoft Vista can handle for other application. However, it requires some steps to complete a task. For example, it requires three steps to thick a checkbox. Therefore, Microsoft Vista speech interaction style is too flexible and not appropriates to apply in the Photo Annotation system, PhotoASYs. Single mode applied in proposed speech interaction style gives users more subjective satisfaction.

## 6 CONCLUSION

This study focuses on empirical evaluation which involves several variables for Web-Based photo annotation system using multimodal interaction (speech interaction). The evaluation involves time completion and subjective satisfaction for both styles of interaction; proposed speech interaction and Microsoft Vista Speech interaction. The evaluation session includes thirteen participants where all of them do not have any experience in using speech recognition on web-based system. They have completed the experiment and answered all questions after the experiment has successfully completed. The screen recording software was used to collect all data. Overall results show that the Web-Based photo annotation system, PhotoASys is more suitable with proposed speech style of interaction rather than Microsoft Vista Speech interaction. The study shows significant different in statistical between proposed speech interaction and Microsoft Vista Speech interaction style in terms of time taken of completion and subjective satisfaction. Hypotheses 1 (H1) and Hypothesis 2 (H2) are rejected since value of Asymp. Sig. (2-tailed) were 0.001 and 0.000 for both hypotheses in which the significant level is less than 0.05. Therefore, both styles of interactions are different in time taken of completion and subjective satisfaction. On average, the reduction in annotation performance time due to using proposed speech interaction was 82.33% rather than using Microsoft Vista speech interaction. As a conclusion, users could choose either proposed speech or Microsoft Vista speech style of interaction. The difference shows the different performance for applying speech for particular system and general system.

### ACKNOWLEDGMENT

This work was supported by GMM, FSKSM, and funded by MOSTI, Ministry of Science, Technology and Innovation Malaysia. Many thanks to all people involved in the experiments and for sharing ideology and opinion during the writing of this paper.

### REFERENCES

[1] 1. O'Keefe, M., 2004. Online photo service usage continues to grow,                                                                   <

http://www.infotrends.com/public/Content/Press/2004/11.17.2004.html >, [last accessed 26/01/2010 ].

[2] 2. Apple iPhoto, 2008 < http://www.apple.com/ilife/iphoto/>, [accessed 20/01/10]

[3] 3. Rodden, K. and Wood. K. R. How do people manage their digital photographs? Proceedings of the SIGCHI conference on Human factors in computing systems. Ft. Lauderdale, Florida, USA, ACM. 2003.

[4] 4. Picasa Online, 2008, < http://picasa.google.com/>,[accessed 25/06/08]

[5] 5. MyPhotoAlbum,2004,<> http://www.myphotoalbum.com/, [accessed 04/07/08]

[6] 6. Snapfish, 2009, < http://www.snapfish.com/>, [accessed 10/01/10]

[7] 7. Chen, J., T. Tan, et al.. A Method for Photograph Indexing Using Speech Annotation. Advances in Multimedia Information Processing — PCM 2001 : 867-872, 2001..

[8] 8. Chen,J., Tan, T., Mulhem, P., and Kankanhalli, M. An Improved Method for Image using Speech Annotation. submitted to Multimedia Modeling. 2003.

[9] 9. Rohini K. Srihari, Zhongfei Zhang, "Show&Tell: A Semi-Automated Image Annotation System," IEEE MultiMedia, vol. 07, no. 3, pp. 61-71, Jul-Sept, 2000.

[10] 10. Karl, L., Pettey, M., Shneiderman, B. Speech-Activated versus Mouse-Activated Commands for Word Processing Applications: An Empirical Evaluation COMSAT Laboratories 23500 COMSAT Drive Clarksburg, MD 2087. 11-34. 1993

[11] 11. Perkakis, M., and Potamianos, A. The Effect of Input Mode on Inactivity and Interaction Times of Multimodal Systems. . 102-109. 2007

[12] 12. Kustanowitz. J, and Shneiderman. B. Motivating Annotation for Personal Digital Photo Libraries: Lowering While Raising Incentives.2004.

[13] 13. Cui, J., W. Fang, et al. EasyAlbum: an interactive photo annotation system based on face clustering and re-ranking. Proceedings of the SIGCHI conference on Human factors in computing systems. San Jose, California, USA, ACM. 367-376, 2007..

[14] 14. B. Suh, and B.B. Bederson, Semi-Automatic Image Annotation Using Event and Torso Identification, 2004.

[15] 15. Chen,J., Tan, T., Mulhem, P., and Kankanhalli, M. An Improved Method for Image Retrieval using Speech Annotation. submitted to Multimedia Modeling. 2003.

[16] 16. Chen, J., T. Tan, et al.. A Method for Photograph Indexing Using Speech Annotation. Advances in Multimedia Information Processing, 867-872, 2001.

[17] 17. K. Christian, B. Kules, B. Shneiderman and A. Youssef, A Comparison of Voice Controlled and Mouse Controlled Web Browsing, proceeding of the fourth in international ACM Conference on Assistive Technologies, pages 72-79, 2000.

[18] 18. N. Alexander, Personal Photo Annotation, September 9, 2005

[19] 19. Kustanowitz. J, and Shneiderman. B. Motivating Annotation for Personal Digital Photo Libraries: Lowering While Raising Incentives.2004.

[20] 20. S.M. Zabed Ahmed, K. Cliff and O. Charles, A user-centered design and evalutiaon of IR interfaces, Journal of Librarianship and Information Science 2006; 38; 157, 2006.

[21] 21. S.A.Ramlan, S.Daut and N.A. Ismail, Group Annotation For Digital Photos Using Multimodal Interaction, The 4th International Conference on Information & Communication Technology and Systems (ICTS) 2008.

[22] 22. S.A.Ramlan and N.A. Ismail, 4w's Framework in Speech Photo Annotation, The 1st International Conference on Research and Innovation in Information System, 2009.

**Siti Azura Ramlan** received an education at Universiti Teknologi Malaysia. She was graduated from Universiti Teknologi Malaysia in Bachelor of Computer Engineering. Now, she active as a researcher and postgraduate student in the Department of Graphic and Multimedia, Faculty of Computer Science and Information System, Universiti Teknologi Malaysia. She has involves in Human Computer Interaction research field especially in speech recognition since 2007.

**Nor Azman Ismail** is an academic staff in Computer Graphics and Multimedia Department, Faculty of Computer Science & Information Systems, Universiti Teknologi Malaysia (UTM). He received his B.Sc. from UTM, Master of Information Technology (MIT) from National University of Malaysia, and Ph.D. in the field of Human Computer Interaction (HCI) from Loughborough University. He has been an academic staff at Computer Graphics and Multimedia Department for about thirteen years and currently, he is Research Coordinator of the Department. Dr. Nor Azman has made various contributions to the field of Human Computer Interaction (HCI) including research, practice, and education.